\documentclass[twocolumn]{aastex63}
\usepackage[utf8]{inputenc}

\usepackage{amsmath}
\usepackage{hyperref}

\graphicspath{{./}{figures/}}

\accepted{October 13, 2019}

\begin{document}

\title{On The Compatibility of Ground-based and Space-based Data: WASP-96 b, An Example}
\correspondingauthor{K.H. Yip}
\email{kai.yip.13@ucl.ac.uk}
\thanks{These authors contributed equally to this work}

\author[0000-0002-9616-1524]{Kai Hou Yip$^*$}
\affiliation{Department of Physics and Astronomy, University College London, Gower Street,WC1E 6BT London, United Kingdom}

\author[0000-0001-6516-4493]{Quentin Changeat$^*$}
\affiliation{Department of Physics and Astronomy, University College London, Gower Street,WC1E 6BT London, United Kingdom}

\author[0000-0002-5494-3237]{Billy Edwards$^*$}
\affiliation{Department of Physics and Astronomy, University College London, Gower Street,WC1E 6BT London, United Kingdom}

\author[0000-0001-8587-2112]{Mario Morvan}
\affiliation{Department of Physics and Astronomy, University College London, Gower Street,WC1E 6BT London, United Kingdom}
	
\author[0000-0002-4552-4559]{Katy L. Chubb }
\affiliation{Department of Physics and Astronomy, University College London, Gower Street,WC1E 6BT London, United Kingdom}
\affiliation{SRON Netherlands Institute for Space Research, Sorbonnelaan 2, 3584 CA, Utrecht, Netherlands}		
 
\author[0000-0003-3840-1793]{Angelos Tsiaras}
\affiliation{Department of Physics and Astronomy, University College London, Gower Street,WC1E 6BT London, United Kingdom}

\author[0000-0002-4205-5267]{Ingo P. Waldmann }
\affiliation{Department of Physics and Astronomy, University College London, Gower Street,WC1E 6BT London, United Kingdom}

\author[0000-0001-6058-6654]{Giovanna Tinetti }
\affiliation{Department of Physics and Astronomy, University College London, Gower Street,WC1E 6BT London, United Kingdom}

\begin{abstract}

    The study of exoplanetary atmospheres relies on detecting minute changes in the transit depth at different wavelengths. To date, a number of ground and space based instruments have been used to obtain transmission spectra of exoplanets in different spectral band. One common practice is to combine observations from different instruments in order to achieve a broader wavelength coverage. We present here two inconsistent observations of WASP-96 b, one by Hubble Space Telescope (HST) and the other by the Very Large Telescope (VLT). We present two key findings in our investigation: 1.) a strong water signature is detected via the HST WFC3 observations. 2.) A notable offset in transit depth ($>1100$ ppm) can be seen when the ground-based and space-based observations are combined together. The discrepancy raises the question of whether observations from different instruments could indeed be combined together. We attempt to align the observations by including an additional parameter in our retrieval studies but are unable to definitively ascertain that the aligned observations are indeed compatible. The case of WASP-96 b signals that compatibility of instruments should not be assumed. While wavelength overlaps between instruments can help, it should be noted that combining datasets remains a risky business. The difficulty in combining observations also strengthens the need for next generation instruments which will possess broader spectral coverage.
    
   \vspace{10mm} 
    
\end{abstract}
\section{Introduction}

The field of exoplanetary science is rapidly expanding, with the discovery of new planets on a weekly basis becoming commonplace.  There is growing interest in gaining a deeper understanding of these worlds and their atmospheric structure. Pioneering works by numerous teams have detected molecular species, alkali metals and other carbon-bearing species presented in the exoplanetary atmosphere  \citep[e.g.][]{charbonneau2002,vidal2004,tinetti2007,Barman2007,Redfield2008,swain2009,Fossati2010,Linsky2010}. Although many of the pioneering works have been done using space-based instruments due to the absence of atmosphere, in recent years ground-based instruments have also made significant contributions to our understanding of exoplanetary atmosphere \citep[e.g.][]{Macintosh2015,Barman2015,Lacour2019,ehrenreich_wasp76,Merritt2020,Bourrier2020}.
    
The installation of Wide-Field-Camera 3 (WFC3) on board of Hubble Space Telescope (HST) has further enriched our understanding of these remote worlds. The introduction of the scanning mode was pivotal in providing high confidence detection of numerous molecular species such as H$_2$O \citep[e.g.][]{ mandell2013, Ehrenreich2014, Wakeford2018,tsiaras_water, Mikal-Evans2020, pluriel_aresIII}, NH$_3$ \citep{Macdonald2017} and TiO \citep{Haynes_Wasp33b_spectrum_em, edwards_w76}. Over the past decade HST WFC3, and other instruments such as Spitzer IRAC and HST STIS, have observed tens of exoplanets and the rich amount of spectral data has led to initial population studies between atmospheres of different exoplanets \citep[e.g.][]{sing2016,iyer_pop,tsiaras_30planets,fisher_18,pinhas}.

WASP-96 b is a transiting gaseous hot-Jupiter discovered by \citet{hellier_wasp96} during the WASP-South Survey. It orbits around a G8 star with a V magnitude of 12.2. Table \ref{tab:orbital_param} summarises the stellar and planetary parameters of the WASP-96 system. WASP-96 b has been observed previously with \citet{Nikolov2018} using FORS2 spectrograph on the Very Large Telescope (VLT). They obtained an optical transmission spectrum spanning from 0.35-0.80 $\mu m$. Their analysis showed that the atmosphere of the planet is cloud-free, under the assumption of chemical equilibrium, and measured an absolute sodium abundance of log $\varepsilon_{Na}$= $6.9^{+0.6}_{-0.4}$ based on the strong sodium profile in the optical waveband. The temperature of the atmosphere was found to be at T = 1710 $^{+150}_{-200}$ K, which is notably higher than the equilibrium temperature of the planet T$_{eq}$ = 1285 $\pm$ 40K. 

In this work we present the HST transmission spectrum of WASP-96\,b, obtained with both the G102 (0.8-1.1 $\mu m$) and G141 (1.1-1.7 $\mu m$) grisms. Our atmospheric retrieval of this data uncovers a strong water signature. To have a more comprehensive analysis of the planetary atmosphere, we attempt to combine this with the data from \citet{Nikolov2018}. However, we find a large offset between the ground-based and space-based datasets. We explore a method of correcting for this issue when there is wavelength overlap and explore the risks associated with combining datasets which cannot be verified to be compatible in absolute transit depth.

%This planet is one of the few planets that were observed by both space based (HST) and ground based instruments (VLT). In this work we are going to present our analysis on this planet using data obtained from G102 and G141 grisms via WFC3. The two grisms have a combined wavelength range from 0.8$\mu m$ to 1.7$\mu m$, which is complementary to previous observation done by \citet{Nikolov2018}. Combining both instruments together will hence provide a continuum from 0.35$\mu m$ (optical) all the way to 1.7$\mu m$ (near-infrared). Given this complementary setup between the two datasets, we will explore the possibility of combining both instruments together and discuss whether combining observations could improve our understanding on the planet.

\section{Data Analysis and Atmospheric Modelling}
\subsection{HST Data Reduction}

The HST data of WASP-96 b were acquired by proposal 15469 led by Nikolay Nikolov and were taken in December 2018. We obtained the raw spatially scanned spectroscopic images from the Mikulski Archive for Space Telescopes\footnote{\url{https://archive.stsci.edu/hst/}} and used Iraclis\footnote{\url{https://github.com/ucl-exoplanets/Iraclis}}, a specialised, open-source software for the analysis of WFC3 scanning observations \citep{tsiaras_hd209}. The reduction process included the following steps: zero-read subtraction, reference pixels correction, non-linearity correction, dark current subtraction, gain conversion, sky background subtraction, calibration, flat-field correction, and corrections for bad pixels and cosmic rays. For a detailed description of these steps, we refer the reader to \cite{tsiaras_hd209}.

\begin{table}
\centering
\begin{tabular}{cc} \hline \hline
Parameters & Value \\ \hline
R$_{s}$ [R$_\oplus$] &$ 1.05 \pm 0.05^*$\\
M$_{s}$ [M$_\oplus$] &$ 1.06 \pm 0.09^*$\\
T$_{s}$ [K] &$ 5540 \pm 140 $\\
M$_p$ [M$_{\rm Jup}$] & $0.48 \pm 0.03^*$ \\
R$_p$ [R$_{\rm Jup}$] & $1.20 \pm 0.06^*$ \\
T$_{\rm Eff}$ (K) &$ 1285 \pm 40^* $\\
a/R$_s$ & $ 8.84 \pm 0.1^\dagger $\\
$i$ [deg] & $85.14 \pm 0.2^\dagger$ \\
P$_{orb}$ [days] & $3.425 2602 \pm 0.000 0027^*$ \\
T$_{mid}$ [BJD$_{TDB}$] & $2456258.062876 \pm 0.0002^*$ \\\hline 
$^*$\citet{hellier_wasp96} & $^\dagger$\citet{Nikolov2018}\\ \hline \hline
\end{tabular}
\caption{Details of the WASP-96 system used in this study.}. 
\label{tab:orbital_param}
\end{table}

\begin{figure}
    \centering
    \includegraphics[width = \columnwidth]{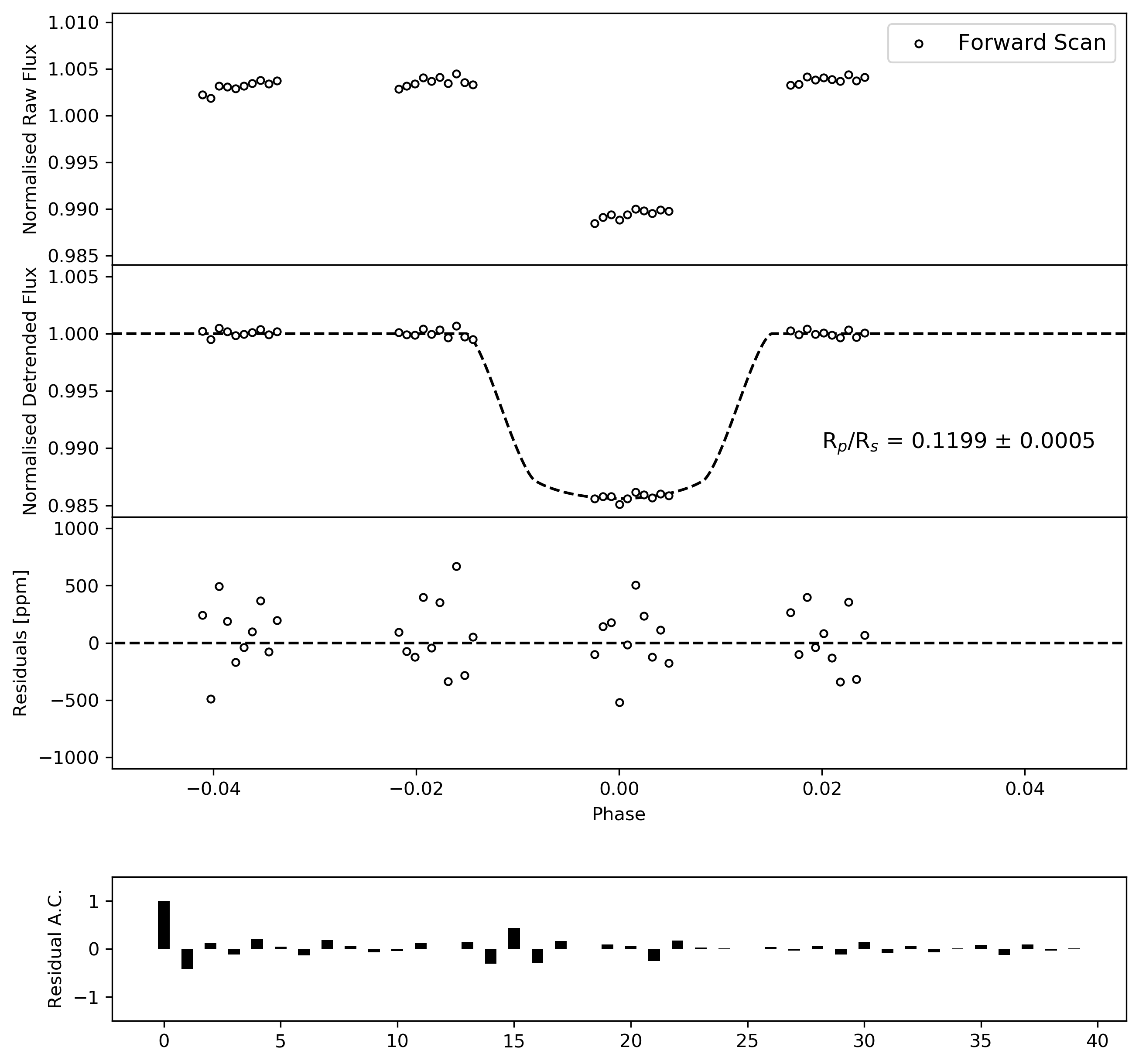}
    \includegraphics[width = \columnwidth]{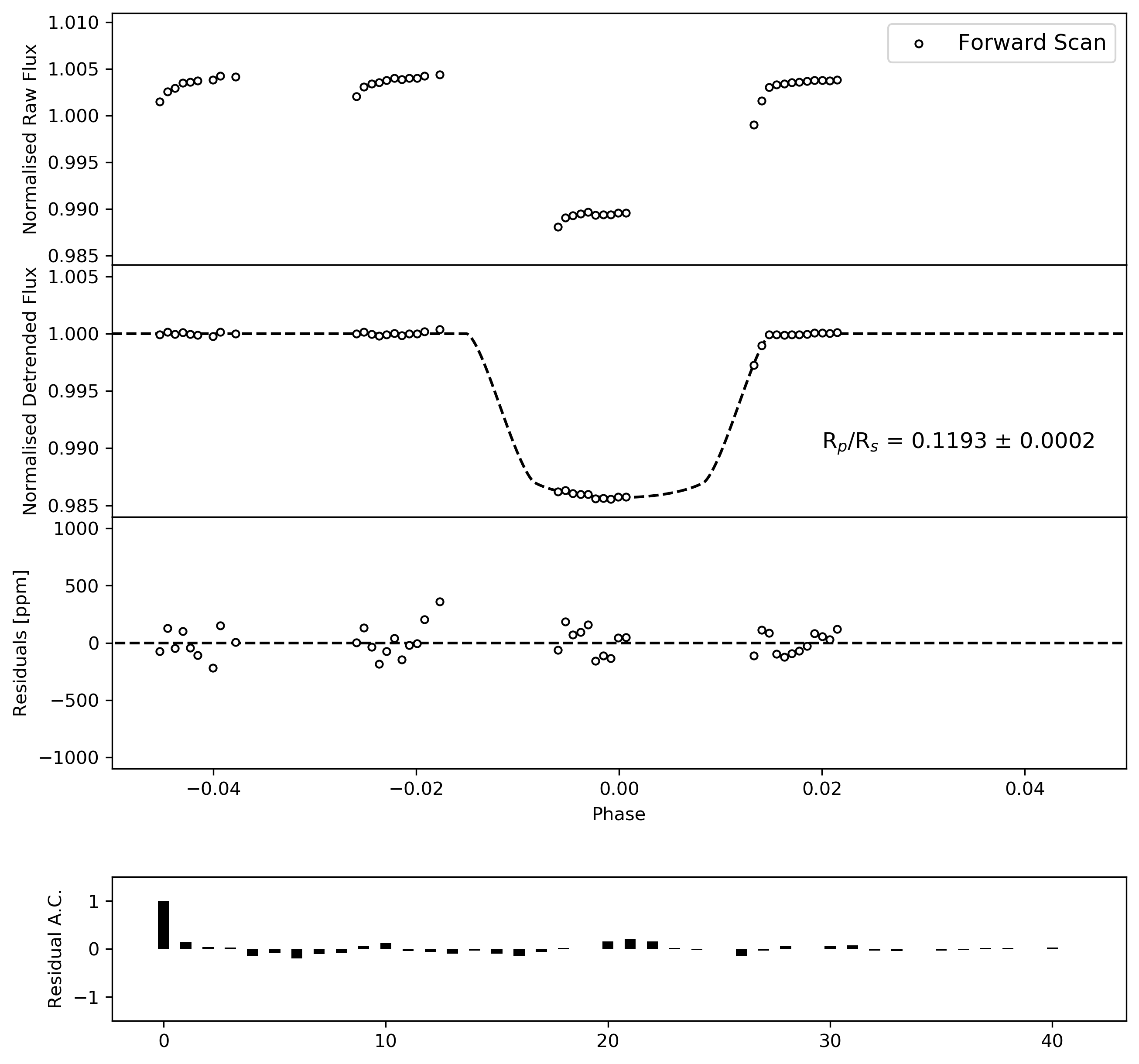}
    \caption{White light curves for the G102 (top) and G141 (bottom) transit observations of WASP-96\,b. First panel: raw light curve, after normalisation. Second panel: light curve, divided by the best fit model for the systematics. Third panel: residuals for best-fit model. Fourth panel: auto-correlation function of the residuals.}
    \label{fig:white_lc}
\end{figure}

\begin{figure*}
    \centering
    \includegraphics[width = \columnwidth]{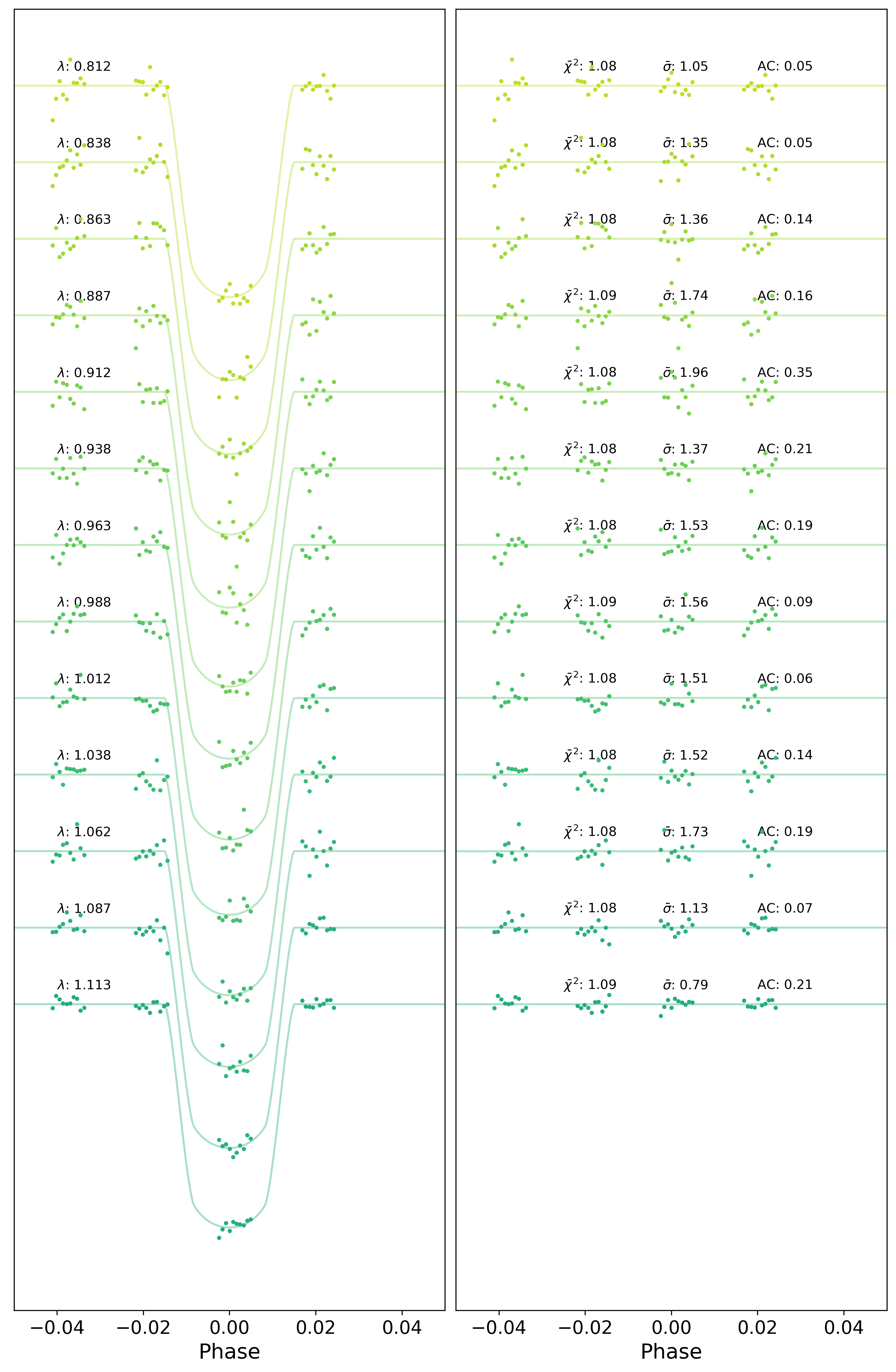}
    \includegraphics[width = \columnwidth]{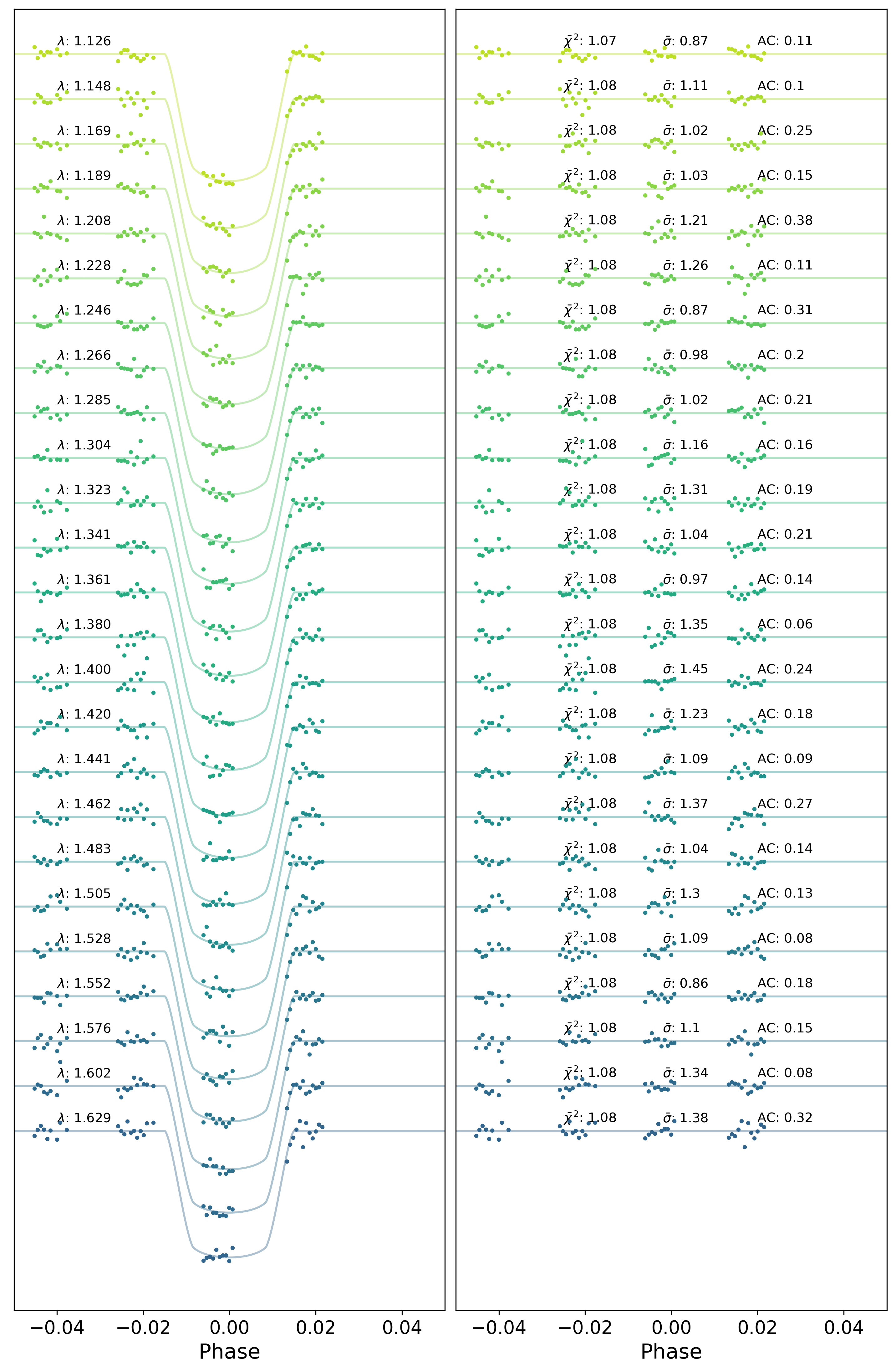}
    \caption{Spectral light curve fits from Iraclis for the G102 (left) and G141 (right) transmission spectra where, for clarity, an offset has been applied. In each plot, left panel: the detrended spectral light curves with best-fit model plotted; right panel: residuals from the fitting with values for the Chi-squared ($\chi^2$), the standard deviation with respect to the photon noise ($\bar{\sigma}$) and the auto-correlation (AC). \vspace{8mm}}
    \label{fig:hst_lc}
\end{figure*}

The reduced spatially scanned spectroscopic images were then used to extract the white and spectral light curves. We then discarded the first orbit of the visit as it presents stronger wavelength dependant ramps. For the fitting of the white light curves, the only free parameters were the mid-transit time and planet-to-star ratio. We did not fit for the inclination or reduced semi-major axis as ingress and egress were not observed in each dataset. The limb-darkening coefficients were selected from using the models of \citet{claretI,claretII} and using the stellar parameters from \cite{hellier_wasp96}. The fitted white and spectral light curves for the G102 and G141 transmission observations are shown in Figures \ref{fig:white_lc} and \ref{fig:hst_lc}.

\subsection{Spitzer IRAC Data Reduction}
\begin{figure}
    \centering
    \includegraphics[width = \columnwidth]{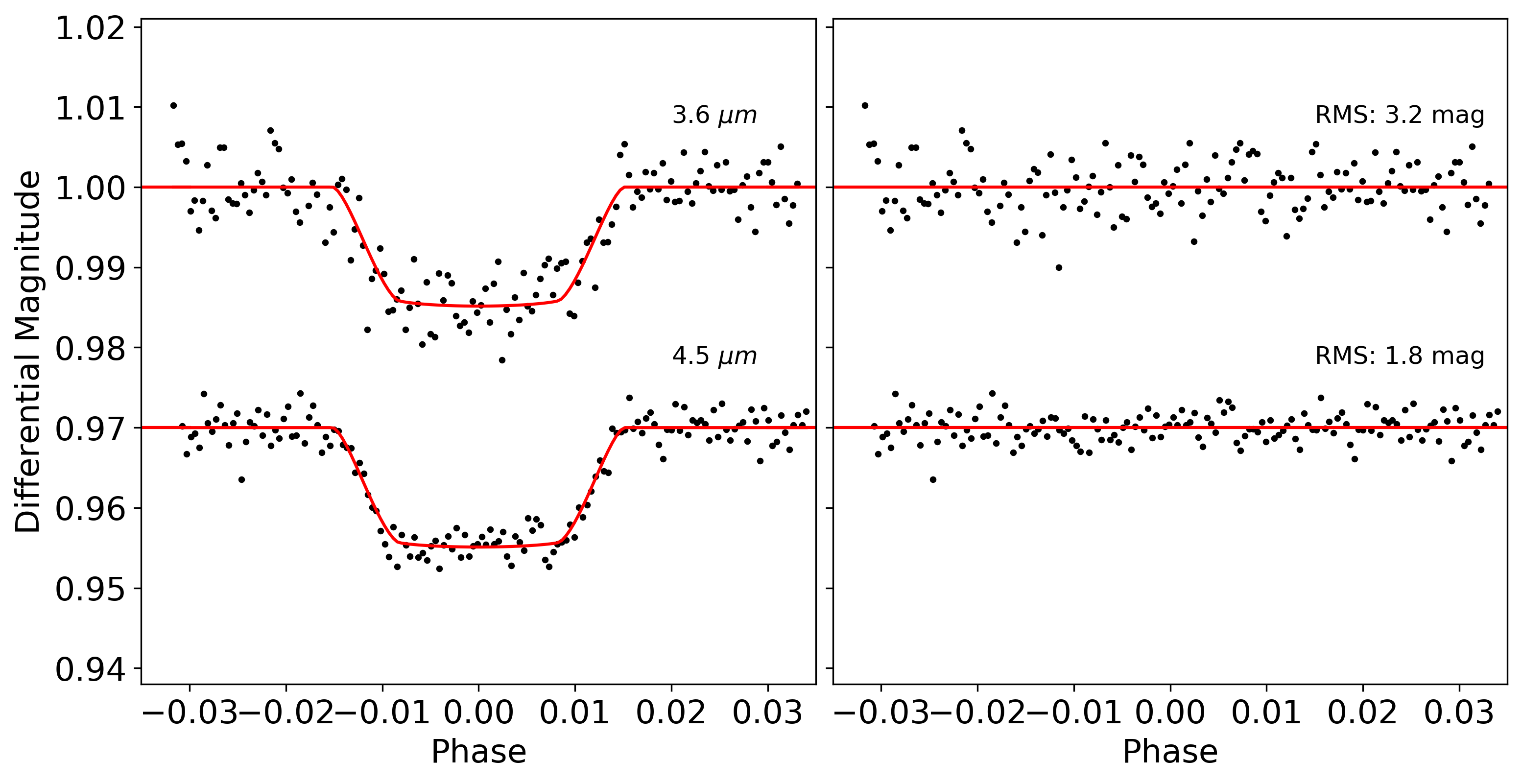}
    \caption{Fitted Spitzer transits of WASP-96 b for each IRAC channel. Left: detrended light curves and best-fit model. Right: residuals from the fitting. \vspace{7mm}}
    \label{fig:spitzer_lc}
\end{figure}

Additionally, two transits of WASP-96\,b had been observed with Spitzer IRAC (program ID 14255). We used the Transit Light Curve Detrending Long Short-Term Memory (TLCD-LSTM) pipeline from \cite{its_a_me_mario} to detrend and fit the Spitzer data from the outer transit and centroids movements. The architecture is the same as in the aforementioned study except for the number of hidden units reduced to 64 and dropout rate set to 0.1 in order to prevent over fitting on the out-of-transit. Figure \ref{fig:spitzer_lc} shows the detrended light curves and the best-fit model to the data. For both datasets, the only free transit parameters were the planet-to-star radius ratio and the transit mid time, with the other model parameters fixed to those in Table \ref{tab:orbital_param}. 

We also fitted the detrended light curve while allowing the planet semi-major axis to star radius ratio (a/R$_{\rm s}$) and inclination (i) to vary. For both channels, the retrieved values for the transit depth and epoch remain very close to the ones found with the two orbital parameters fixed to the values from \citet{Nikolov2018}. Furthermore, values retrieved for the semi-major axis and inclination shown in Table  \ref{tab:orb_fit} are compatible with those from \citet{Nikolov2018} to 1$\sigma$.

{}

\subsection{TESS Data Reduction}

\begin{figure}
    \centering
    \includegraphics[width = \columnwidth]{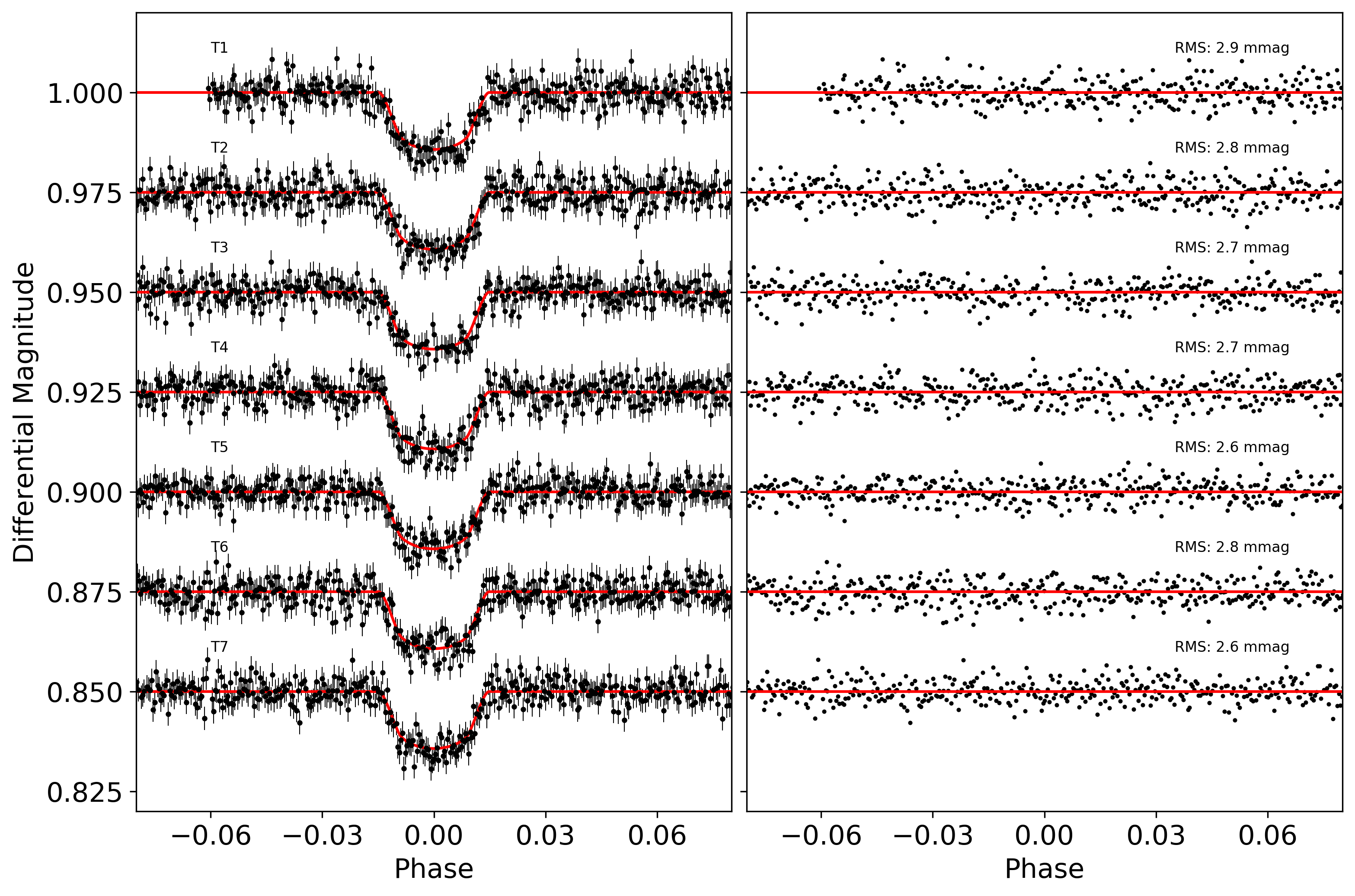}
    \caption{Fitted TESS transits of WASP-96 b. Left: detrended light curves and best-fit model. Right: residuals from the fitting.}
    \label{fig:tess_lc}
\end{figure}

Keeping exoplanet transit ephemeris fresh is crucial for allowing further atmospheric characterisation. Here, we use our Hubble observations along with data from Spitzer and the Transiting Exoplanet Survey Satellite (TESS, \citet{ricker}) to update the orbital period and transit epoch of WASP-96\,b. TESS data is publicly available through the MAST archive and we follow the procedure from \cite{edwards_orbyts} to download, clean and fit the 2 minute cadence Pre-search Data Conditioning (PDC) light curves \citep{smith_pdc,stumpe_pdc1,stumpe_pdc2}. WASP-96\,b was observed during Sector 2 and, after excluding bad data, 7 transits were recovered. These were individually fit and are shown in Figure \ref{fig:tess_lc}. In our main analysis, the only free parameters were the transit mid time and the planet-to-star radius ratio. However, we also performed a fitting of all the TESS data where further orbital parameters, namely the inclination (i) and planet semi-major axis to star radius ratio (a/R$_{\rm s}$) were allowed to vary. The results are summarised in Table \ref{tab:orb_fit} along with similar fits for the Spitzer data.

\begin{table*}[]
    \centering
    \begin{tabular}{ccccc}\hline \hline
      Parameter &  \citet{Nikolov2018} & TESS & IRAC C1 & IRAC C2 \\\hline \hline
       a/R$_{\rm s}$ & 8.84 $\pm$ 0.1 & 8.85 $^{+0.62}_{-0.10}$ &  8.66 $^{+0.14}_{-0.12}$ & 8.78 $^{+0.06}_{-0.06}$ \\
       i [deg] & 85.14 $\pm$ 0.2 & 85.55 $^{+0.39}_{-0.34}$ & 85.36 $^{+0.14}_{-0.13}$ &  85.21 $^{+0.07}_{-0.06}$\\\hline \hline
    \end{tabular}
    \caption{Comparison of orbital parameters between different datasets.}
    \label{tab:orb_fit}
\end{table*}

\subsection{Atmospheric Modelling}

The retrieval of the transmission spectrum was performed using the publicly available retrieval suite TauREx 3 \citep{al-refaie_taurex3}\footnote{\url{https://github.com/ucl-exoplanets/TauREx3_public}}. For the star parameters and the planet mass, we used the values from \citep{hellier_wasp96}. In our runs we assumed that WASP-96\,b possesses a primary atmosphere with a ratio He/H$_2$ = 0.17 (i.e. solar abundance). To this, we added trace gases and included the molecular opacities from the ExoMol \citep{Tennyson_exomol}, HITRAN \citep{gordon} and HITEMP \citep{rothman} databases for: H$_2$O \citep{polyansky_h2o}, CH$_4$ \citep{jt698}, CO \citep{li_co_2015}, CO$_2$ \citep{rothman_hitremp_2010} NH$_3$ \citep{jt771}, K and Na \citep{NISTWebsite}. The line broadened profiles for the resonance doublets of Na and K are computed using \cite{16AlSpKi.broad} and \cite{19AlSpLe.broad}. For the clouds, we use grey opaque clouds and the Mie cloud model from \cite{Lee_haze_model}. On top of this, we also included Collision Induced Absorption (CIA) from H$_2$-H$_2$ \citep{abel_h2-h2, fletcher_h2-h2} and H$_2$-He \citep{abel_h2-he} as well as Rayleigh scattering for all molecules. We assumed iso-thermal and iso-chemical profiles throughout all our retrievals.

In our retrieval analysis, we used log uniform priors for all parameters as described in Table \ref{tab:priors}. Finally, we explored the parameter space using the nested sampling algorithm Multinest \citep{Feroz_multinest} with 750 live points and an evidence tolerance of 0.5.

\begin{table}
    \centering
    \begin{tabular}{cccc} \hline  \hline
    \multicolumn{4}{c}{Priors}\\ \hline
    Parameters & \multicolumn{2}{c}{Prior bounds} & Scale\\ \hline 
    H$_2$O & -12 & -2 & log\\
    CH$_4$ & -12 & -2 & log\\
    CO & -12 & -2 & log\\
    CO$_2$ & -12 & -2 & log \\
    NH$_3$ & -12 & -2 & log \\
    $T_{term}$ (K)& 400 & 2000 & linear \\
    $P_{clouds}$ (Pa)& 6 & 1 & log  \\
    $R_p$ (R$_{jup}$)& 0.6 & 2.4 & linear \\ \hline
    K & -12 & -2 & log\\
    Na & -12 & -2 & log\\
    $P_{mie}$ (Pa) & 6 & 0 & log  \\
    $\chi_{mie}$ & -20 & -5 & log  \\
    Offset [ppm] & -5000 & 5000 & linear  \\\hline \hline
    \end{tabular}
    \caption{List of the retrieved parameters, their uniform prior bounds and the scaling used. The top half of the table shows the parameters fitted for during the HST only retrieval and the lower half contains those that were during the fitting which included the VLT data.}
    \label{tab:priors}
\end{table}

\begin{figure*}
    \centering
    \includegraphics[width = \textwidth]{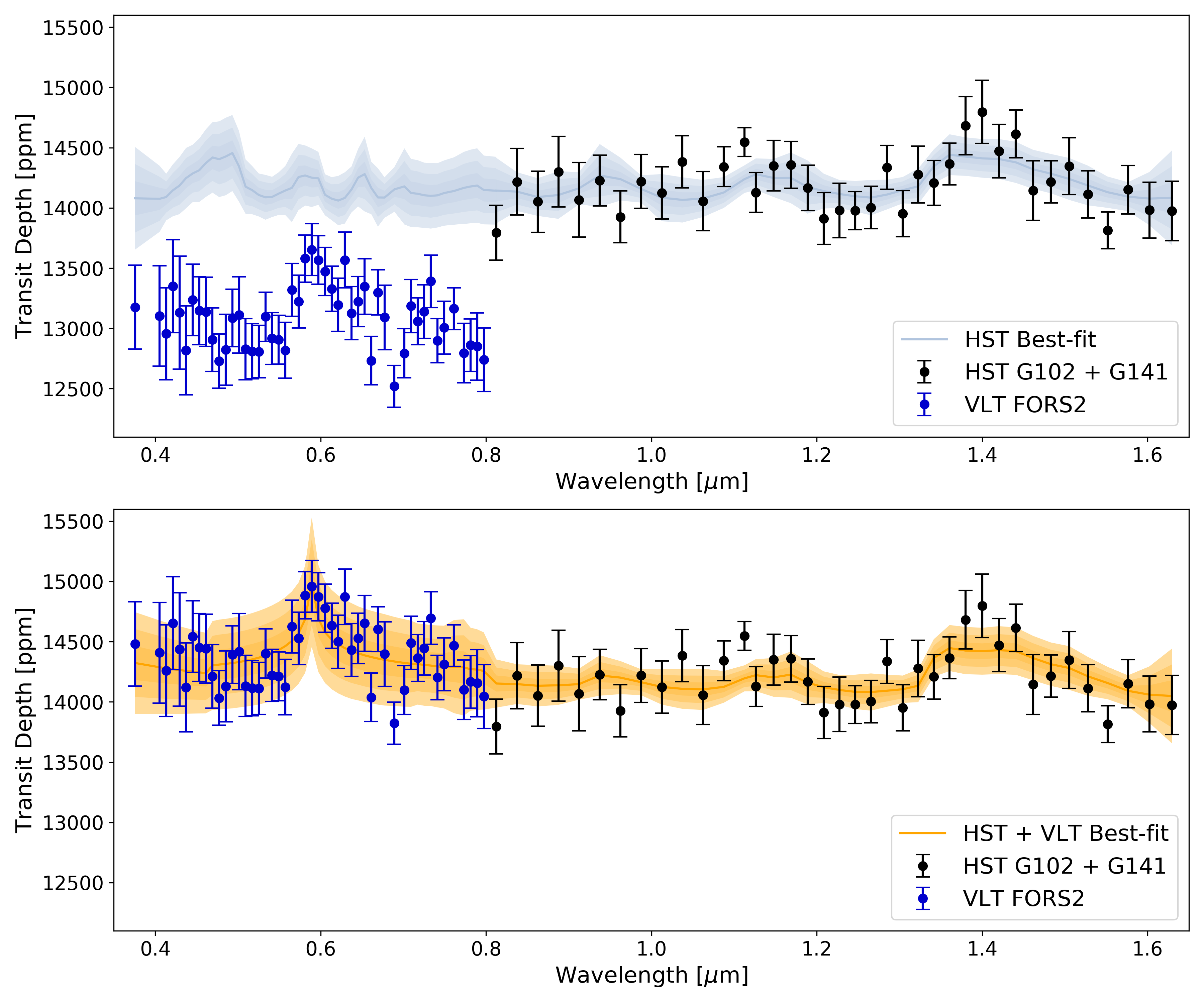}
    \caption{Transit depths derived from HST data (this work) with the addition of data from the VLT \citep{Nikolov2018}. A clear offset can be seen between the ground-based and space-based datasets (top panel). In both panels we overplot the best fit spectrum for HST only (top, in blue) and HST+VLT after correction (bottom, in orange).} %Additionally we plot the TESS data for reference and, although we did not include it in the retrieval analysis, it agrees with the transit depths derived from HST.}
    \label{fig:all_obs}
\end{figure*}{}

\section{Results}
\subsection{Retrieval results: HST observation only}
\label{sec:HST_retri}

\begin{figure*}
    \centering
    \includegraphics[width = \textwidth]{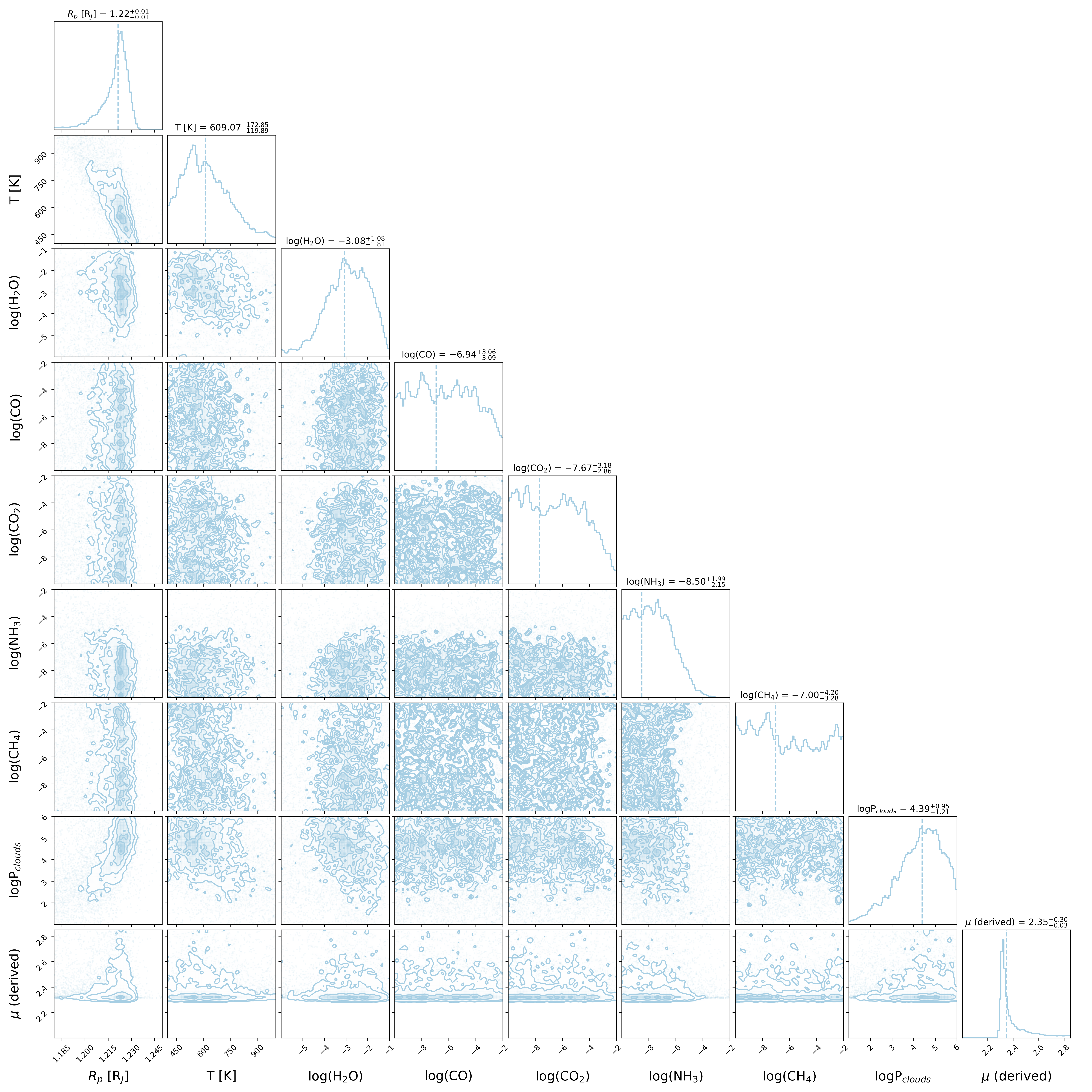}
    \caption{Posteriors distribution of different atmospheric parameters retrieved using HST data only (top of Figure \ref{fig:all_obs}.)  }
    \label{fig:hst_posteriors}
\end{figure*}{}

The top panel of Figure \ref{fig:all_obs} shows the best-fit spectrum from our retrieval based on G102 and G141 data only.  The best-fit model contains modulations which correspond to multiple absorption features of water, clearly indicating the presence of the molecule in the atmosphere. Our free Bayesian retrieval analysis recovered a water abundance of log(H$_2$O) = -3.08$ ^{+1.08}_{-1.81}$, consistent with predictions from equilibrium chemistry models \citep{Agundez_2012_ACE, Woitke_2018, Stock_2018}. Assuming an isothermal temperature profile, our retrieval determines a temperature T = 609$^{+173}_{-120}$K, which is lower than the equilibrium temperature of the planet and much lower than the temperature obtained by \citet{Nikolov2018}. Simulations have demonstrated that the complexity of the terminator limb, which includes 3-dimensional asymmetries in the chemical and thermal structure of the terminator region, often affects the absolute retrieved temperature and can have biases \citep{caldas, pluriel_bias, macdonald_cold}. A lower than expected temperature is often retrieved from HST WFC3 data \citep[e.g.][]{skaf2020}. Similarly, narrow wavelength coverage can result in wrong estimations of the atmospheric temperatures, which can be unstable if a single molecular band is probed \citep[e.g.][]{rocchetto,tsiaras_30planets, pinhas}. On the other hand, our analysis also shows a grey cloud top pressure at log(P$_{clouds}$) = $4.39^{+0.95}_{-1.21}$, which seems to suggest a relatively clear atmosphere, confirming the findings of \cite{Nikolov2018}. We did not fit for more complicated scattering models as the HST wavelength range does not cover a sufficiently large wavelength range (in particular shorter wavelength) to constrain atmospheric scattering model. (see Appendix \ref{fig:HST_complex_posterior} for results from fitting more complicated scattering model.). Due to the weak absorptions, especially compared to water, of other molecules in the wavelengths considered, we were not able to determine the molecular abundance of NH$_3$, CO and CH$_4$. For NH$_3$, we were able to extract a 1$\sigma$ upper bound of log(NH$_3$)$_{upper}$ = -6.51. The posterior distributions for this retrieval are shown in Figure \ref{fig:hst_posteriors}.

\subsection{Offset between HST and VLT observations}

The top panel of Figure \ref{fig:all_obs} shows the HST spectra (G102 and G141) analysed here. We also display the raw spectrum obtained with the VLT in \citep{Nikolov2018}. With no correction, we can immediately observe that the two sets of spectra are not compatible. In particular, a significant offset between the the ground-based (orange) and space-based data (blue and dark blue) at around 0.8 $\mu$m. There are a number of potential sources for the differences seen here: variations in the stellar properties, instrument systematics, differences in the reduction pipelines, telluric corrections and the use of different orbital parameters or limb darkening coefficients. 

\begin{figure}
    \centering
    \includegraphics[width = \columnwidth]{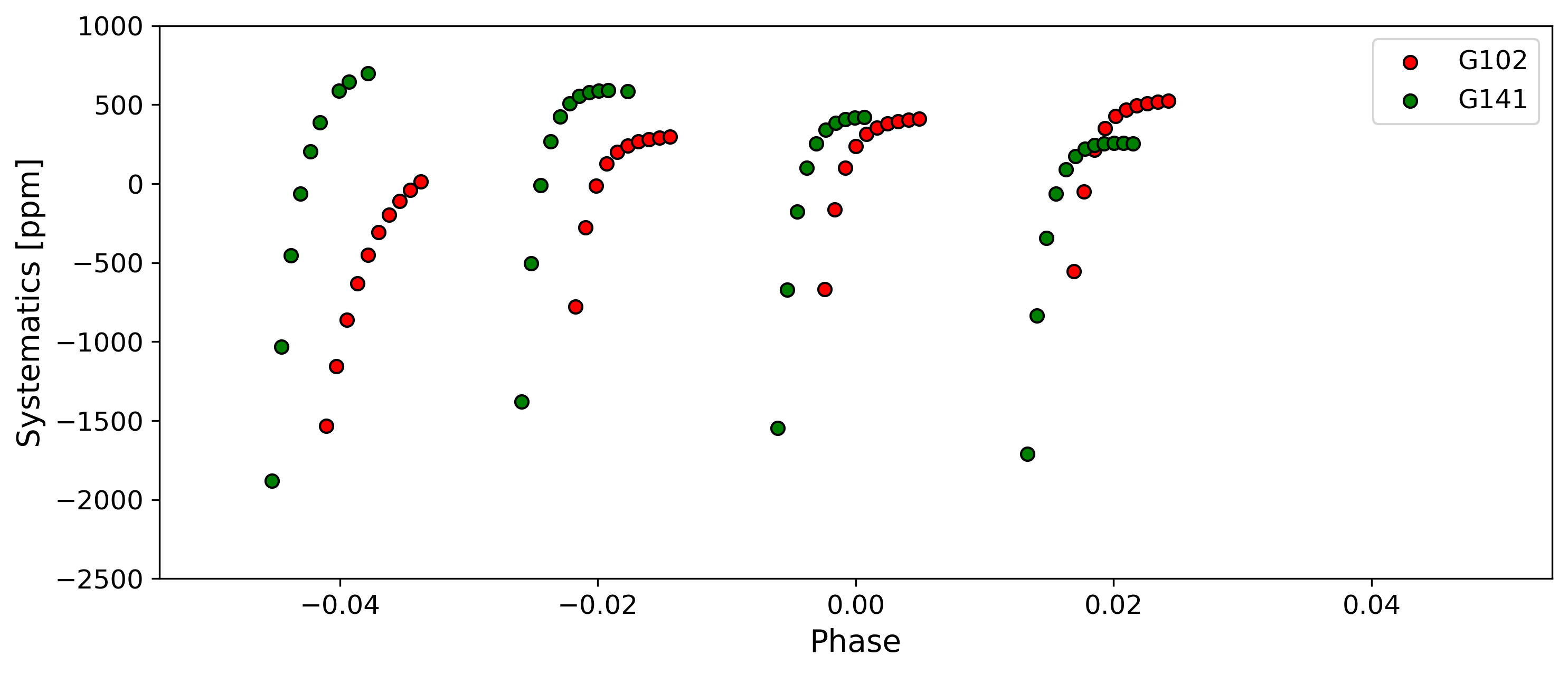}
    \caption{Best-fit systematics for the two HST observations analysed here}
    \label{fig:systematics}
\end{figure}

An imperfect correction of instrument systematics certainly has the potential to significantly alter the recovered transit depth, the best-fit models of the systematics, namely the orbital and long period ramps, are shown in Figure \ref{fig:systematics}. The corrections applied to some exposures are greater than the offset seen between the HST and VLT data, which may explain the offset observed. However, the ramps seen in the G102 and G141 data are very different yet they have both been fitted such that the final data products are seemingly in good agreement. The observations had different exposure times (179.05 s; 156.70 s), scan lengths (2.41 "; 3.60 ") and scan rates (0.013 "/s; 0.022 "/s) so different systematics are to be expected.

For both HST observations we fitted a linear long-term trend, in line with many previous studies.  \citet{guo_hd97658} suspected, from visual inspection, that the trends seen in the WFC3 data of HD\,97658\,b deviated from a linear fit. Thus, they experimented with a number of different trends (quadratic, exponential, logarithmic) finding that, while the lightcurve depths recovered from each observation were roughly consistent with one another, the chosen trend affected the transit depth recovered from the white light curves. They noted that the quadratic model produced the most uncertain depth as well as being the most discrepant between visits. \citet{agol_spitzer} also saw a bias when fitting Spitzer data with a quadratic trend. Here, visual examination of the raw light curves does not indicate that long term trends are non-linear and thus we do not explore different detrending modules.

In \citet{Nikolov2018}, the authors noted that the derived depths from their two visits had marginal disagreement (1.4$\sigma$). Said difference was 720 ppm but the authors noted that this level of variation is consistent with the photometric variability of the star, which is associated with active regions on its surface, of 920 ppm. The variation in the stellar flux could therefore be the cause of much of the offset seen here. We note that this could also be causing differences between the G102 and G141 observations though the derived depths are within 1$\sigma$. These were taken 10 days apart while the VLT datasets were acquired with a gap of 24 days.

The effect of combining different instruments is studied in the literature. \citet{alexoudi_inc} highlighted the importance of the using the correct orbital parameters for data covering visible wavelengths: otherwise slopes can be induced, or removed. \citet{yip} highlight the danger of combining data from HST and the Spitzer Space Telescope, which are not sharing a common baseline and where the information redundancy for carbon based species is limited. Despite this, studies often combine these data \citep[e.g.][]{sing2016, pinhas}. The combination of individual analyses from two different sets of observations may not necessarily agree with each other and should be approached with care. 

The case we presented here is an obvious example of when the offset can be visually inspected but it is not the first study to introduce an offset to a dataset. \citet{giovanni2020} presented that, for the active star WASP-52, stellar spots could create incoherent observations. The team combined transit data from HST/STIS, HST/WFC3, and Spitzer/IRAC of WASP-52\,b and corrected for the offset in HST/WFC3 by accounting for the effect of stellar spots before retrieving the planet's atmospheric composition based on the corrected observations. 

\citet{kirk_w39} explored the effect of stellar activity on the atmospheric retrieval of WASP-39\,b when combining  different datasets. In that study, the infrared data came from HST WFC3 G102 and G141 while the optical datasets were from the ground-based ACAM instrument on the 4.2 m William Herschel Telescope or from HST STIS, providing continuous coverage from 0.4-1.6 $\mu m$. They found no noticeable offset and little difference in the retrieved parameters when accounting for stellar activity, suggesting that, for WASP-39\,b, the datasets could be compatible. 

Other studies have applied, or fitted for, offsets without wavelength overlap. In a study of WASP-74\,b with photometry from ground-based instruments along with HST/WFC3 and Spitzer/IRAC, \citet{luque_w74} fitted for an offset for the HST data, discovering a best-fit value of 434 or 615 ppm, depending upon the additional data used. \citet{wilson_w103} obtained FORS2 data of WASP-103\,b and combined it with other ground-based data as well as data from HST and Spitzer. They applied a small offset to match the GMOS and FORS data before including an offset parameter in their retrievals to account for any further discrepancies. Meanwhile, in their study of HAT-P-12\,b, \citet{yan_h12} attempted to fit for two offsets to improve their fitting of HST WFC3, HST STIS and LBT data but found it did not led to solutions which were statistically more valid. Finally, when fitting for an offset in the case of WASP-69\,b between HST WFC3 and OSIRIS observations, a value of 479 or 618 ppm was recovered by \citet{murgas_offset} with/without also fitting for a spot correction. They noted the cause could be biases due to instrument systematics but that largest semi-amplitude of the WASP photometry, 13 mmag, would result in a flux variation of 2.4\% which could account for the 2.2\% increase in flux between the observations.

However, there have been other instances in the literature where ground-based and space-based observations are combined and analysed without wavelength overlap nor offset correction. For example, \cite{Danielski2014} combined observations from NASA Infrared Telescope Facility (IRTF)/SpeX instrument and WFC3 observations on HD 189733b, many others followed similar trends \citep[e.g.][]{mancini_w19, bean_w19, stevenson_h26, sotzen_w79}. While these datasets could be compatible, there is no guarantee and this should not be taken for granted.  

In our case, without assigning a specific cause, we attempted to correct for the offset by fitting a single flat offset parameter in a combined retrieval. The parameter applies a shift to the entire VLT spectrum vertically to create a coherent observation with the HST dataset. This choice of correction does not necessarily represent the complexity of the problem here and it is not guaranteed that the instrument systematics are wavelength independent.

\begin{figure*}
    \centering
    \includegraphics[width = \textwidth]{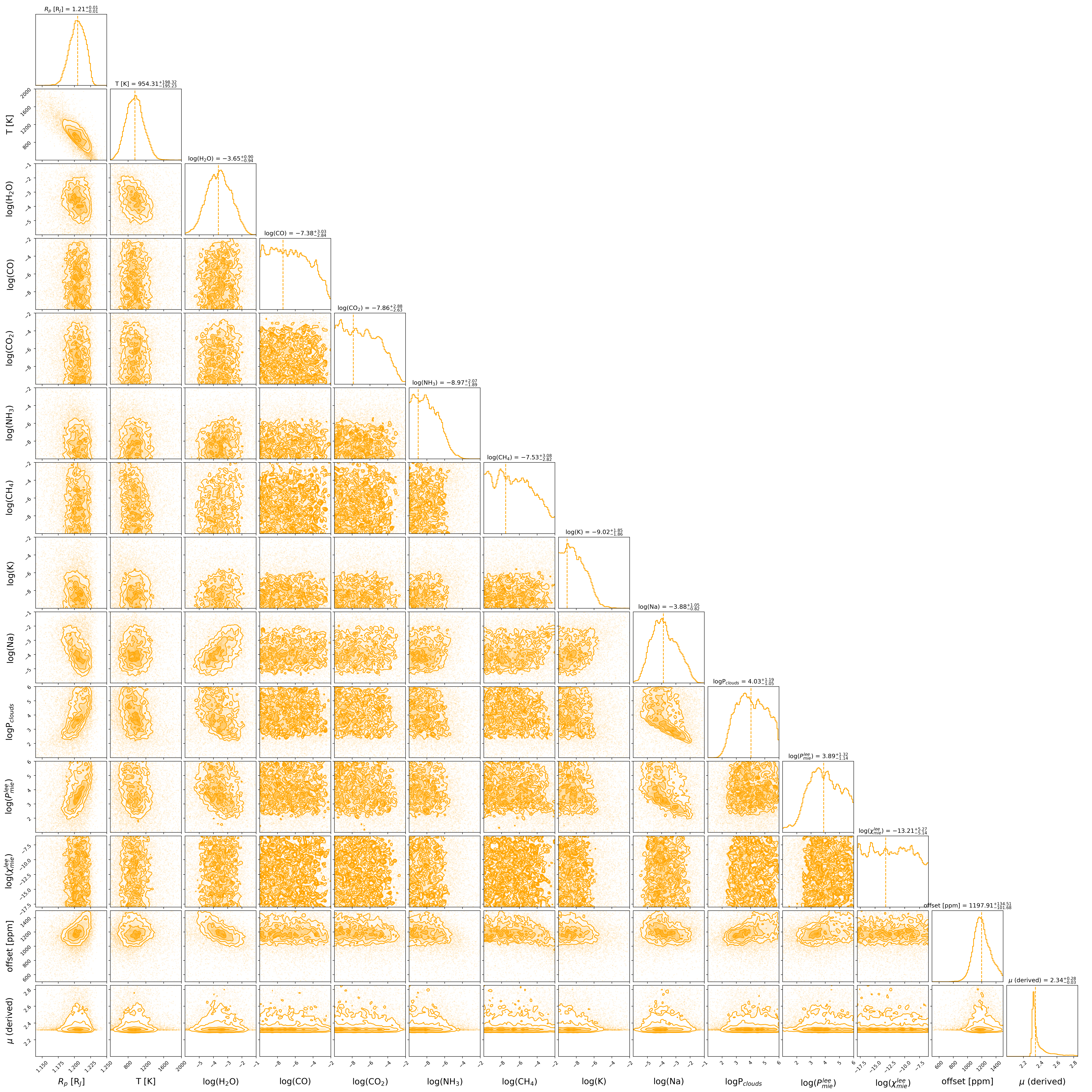}
    \caption{Posteriors distribution of different atmospheric parameters retrieved using both HST data and VLT data (after offset correction, bottom of Figure \ref{fig:all_obs}) }
    \label{fig:offset_posteriors}
\end{figure*}{}

\subsection{Retrievals results: HST + VLT observations}
\label{hst_vlt_obs}
The best-fit spectrum is plotted in the bottom panel of Figure \ref{fig:all_obs}. Our combined retrieval unveiled an offset of 1198 $^{+135}_{-102}$ ppm between the two instruments as shown in the posterior distributions in Figure \ref{fig:offset_posteriors}. The increased wavelength coverage (continuous coverage from 0.4-1.6 $\mu$m) allows us to fit for more complicated cloud models and probe the presence of two species (H$_2$O and Na). The extension to visible also helped to provide better constraint on the temperature on the terminator. The retrieved temperature (T = 954 $^{+198}_{-195}$ K) is close to the expected terminator temperature from a linear trend derived in \citet{skaf2020}, given the equilibrium temperature of the planet. An interesting potential explanation for the large temperatures difference between the 1700 K inferred from the VLT only data and the one retrieved from the HST only data could be that the signal from those two molecules comes from different regions of the terminator. \citet{caldas}, taking the example of H$_2$O and CO, predicted that some molecules could have a signal from the day-side part of the terminator region only or inversely the night-side part only. 

The retrieved abundances for H$_2$O and Na in our combined retrieval are within 2-$\sigma$ agreement with the individual analysis reported in our work and \citet{Nikolov2018} (See Table \ref{tab:all_retrievals} for comparison between retrieval scenarios). The consistency in the water abundances indicates that the water feature may be stable enough to be retrieved accurately with HST, even in the case of the low terminator temperature recovered in Section \ref{sec:HST_retri}. Such a result is expected given the strong features of H$_2$O in the WFC3 range and lack thereof in the visible. 
%\citet{Casasayas2020} showed that the apparent Na signal seen in HD\,209458\,b could be explained by the Rossiter McLaughlin \citep{rossiter,mclaughlin} effect and we note that this could potentially be contributing to the peak seen here.

\begin{table}
    \centering
    \resizebox{\columnwidth}{!}{%
    \begin{tabular}{c|ccc}\hline \hline
    Parameter & HST   & (Nikolov et al.) & HST+VLT \\ \hline
    log(H$_2$O) & $-3.08 ^{+1.08}_{-1.81}$ & N/A & $-3.65 ^{+0.90}_{-0.94}$ \\
    log(Na) & N/A & $-5.1 ^{+0.6}_{-0.4}$ & $-3.88 ^{+1.05}_{-0.82}$ \\
    R$_p$ [R$_{J}$] & $1.22 ^{+0.01}_{-0.01}$ & N/A & $1.21 ^{+0.01}_{-0.01}$ \\
    T [K] & $609^{+173}_{-120}$ & $1710 ^{+150}_{-200}$ & 954 $^{+198}_{-195}$ \\  \hline \hline
    \end{tabular}
    }
    \caption{Comparison between different retrieved quantities for three different scenarios (HST only, VLT only and HST+VLT). Results from VLT alone is reproduced from \citet{Nikolov2018}. We have omitted  quantities that were unable to be constrained by any of the scenarios.  }
    \label{tab:all_retrievals}
\end{table}

Based on the retrieval result and the visually compatible observations in Figure \ref{fig:all_obs}, it may be tempting to conclude that the correction has been successful. However, we would like to emphasise here that this kind of correction is an ad-hoc solution to the problem and does not contain any theoretical support. Any conclusion drawn from this kind of combined observations should be treated lightly and comparisons to model fitting on single datasets made. 

A retrieval study of 10 hot Jupiters by \cite{pinhas} found that optical data played significant role in ensuring that a reliable constraint on the abundances derived from infrared data could be placed. When combining optical and infrared data of HD\,209458\,b from HST (STIS + G141), they found the constraints on water to be narrowed by a factor of 3. However, the abundances retrieved in each case were drastically different: log(H$_2$O)$_{WFC3}$ = -3.3$^{+0.80}_{-0.75}$ and log(H$_2$O)$_{STIS + WFC3}$ = -4.66$^{+0.39}_{-0.30}$. 

In contrast, the water abundances recovered here, with and without optical data, are in good agreement with one another (well within 1$\sigma$). This may well be due to the addition of the G102 grism, which has not been used to observe HD\,209458\,b, highlighting that it has the capability to provide excellent constraints on the water abundance when combined with G141. Additionally, while here we focus on a case where ground-based and space-based instrument demonstrate an offset, such a discrepancy could also occur in space-based datasets from different instruments. Combining HST STIS, HST WFC3 G141 and Spitzer IRAC has become common place in the field \cite[e.g.][]{sing2016}. Given that there is no wavelength overlap in these studies, there is a risk of offsets occurring which could bias the results of subsequent atmospheric retrievals. For instances, the Spitzer IRAC bands cover spectral regions where carbon-bearing molecules such as CH$_{4}$, CO and CO$_2$ absorb, which could be biased when the instrument is not well calibrated, leading to wrong estimates of C/O ratio. The G102 grism remains an under utilised instrument for exoplanet spectroscopy but would offer extra confidence that the STIS and G141 datasets are compatible by providing wavelength overlap with both.

\subsection{TESS and Spitzer Transit Depths}

\begin{figure}
    \centering
    \includegraphics[width=\columnwidth]{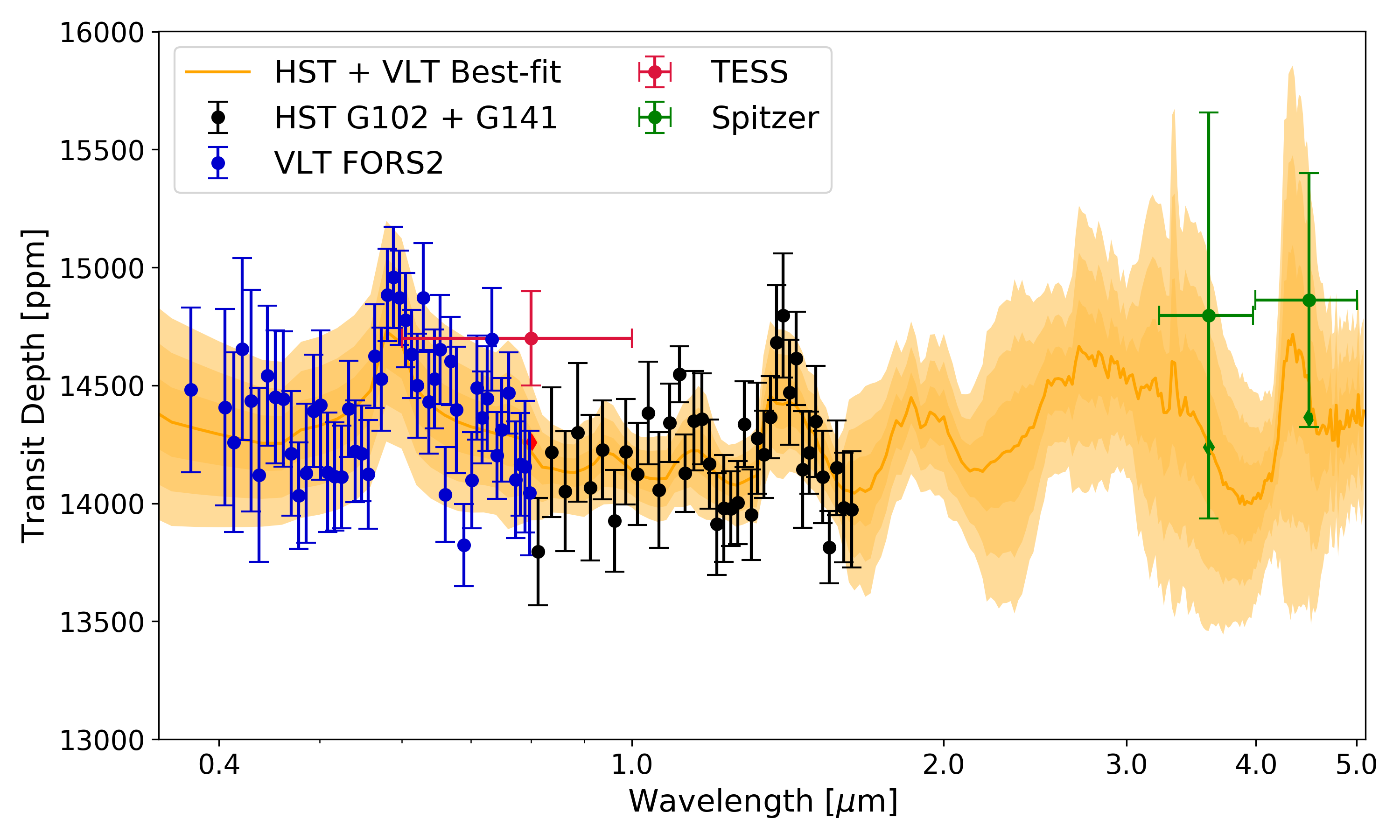}
    \caption{Best-fit model to the VLT and HST observations with data from TESS and Spitzer over-plotted. The diamonds denote the transit depth of the model across these entire bands. The Spitzer data is within 1$\sigma$ of the retrieval model but the error bars are so high that little extra information could be gained by including them in the retrieval. Additionally, the risk of an offset is still present but undetectable given the lack of wavelength overlap.}
    \label{fig:bestfit_spitzer}
\end{figure}

\begin{figure}
    \centering
    \includegraphics[width = \columnwidth]{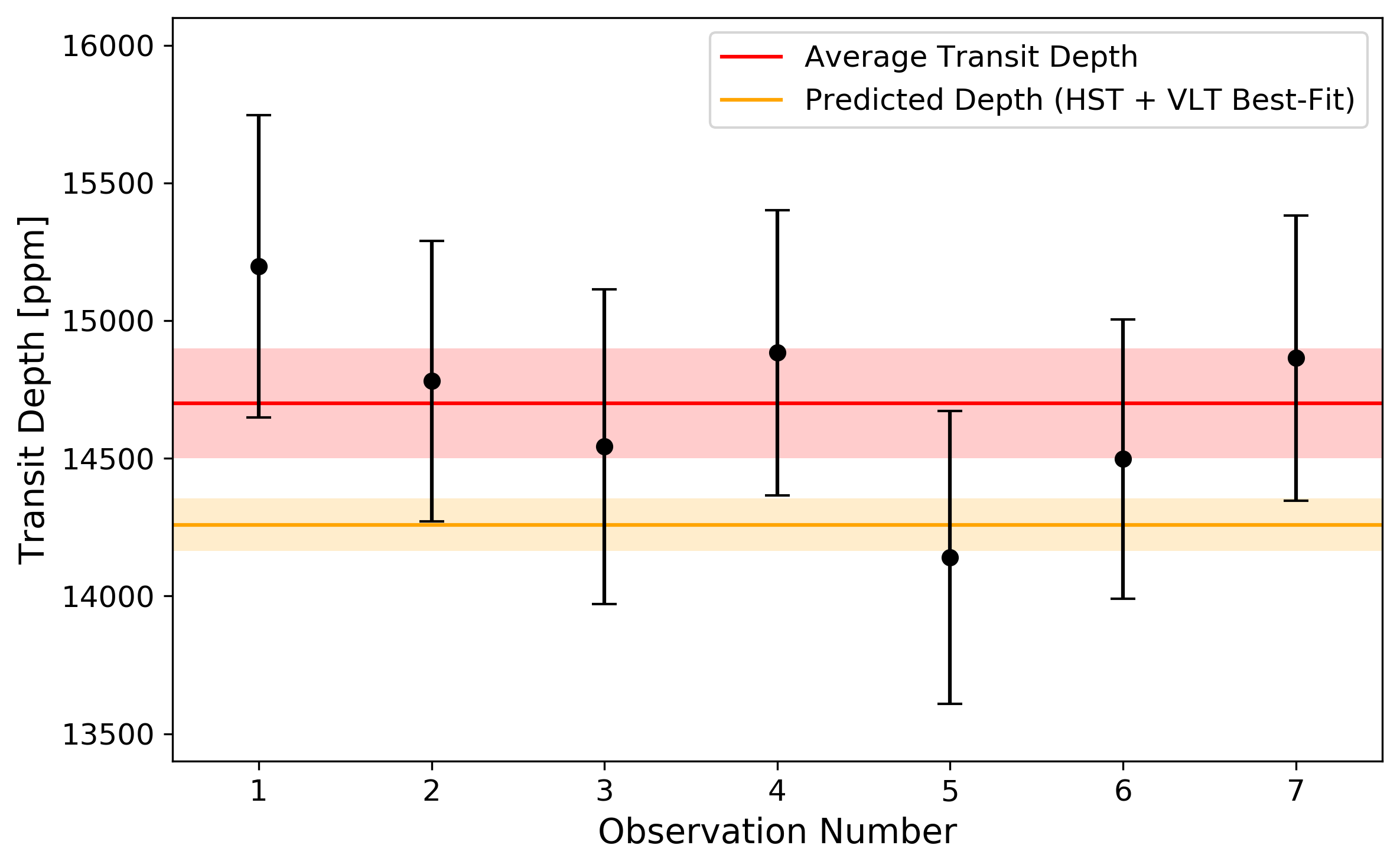}
    \caption{TESS transit depths from each individual observations and their weighted average. This can be seen to disagree by $2\sigma$ compared to the expected depth from our HST + VLT retrieval.}
    \label{fig:tess_depths}
\end{figure}

In addition to VLT observation, we have also explored the scenario when observations from Spitzer and TESS are added to the HST data. Figure \ref{fig:bestfit_spitzer} shows the TESS and Spitzer data plotted alongside the best-fit spectrum from the HST and corrected VLT retrieval from the previous section. The two Spitzer points (3.6 $\mu$m and 4.5 $\mu m$ ) are within 1$\sigma$ of the best fit solution. Hence, it is possible that Spitzer is consistent with the other instruments. However, we remain cautious and, as there is no wavelength overlap for Spitzer, we cannot use the methodology employed for the correction of the VLT data. Additionally, given the size of the error bars on the Spitzer data, little additional spectral information would be gained. The TESS point, on the other hand, is about 2$\sigma$ away from the solution. Seven transits were observed with TESS and the depth of each of these is shown in Figure \ref{fig:tess_depths}. While several of the individual observations are consistent with the HST + VLT best-fit (to $1\sigma$), the weighted mean of these observations is larger than the model (orange bar). As the average transit depth is larger than that obtained with HST, and the Spitzer points are seemingly consistent with the HST data. It provides further indications that the source of the offset seen between datasets may be caused by the reduction and analysis of the VLT observations. The transit depth data for all instruments is given in Table \ref{tab:spec_data}.

\subsection{Ephemeris Refinement}

\begin{table}
    \centering
    \begin{tabular}{ccc}\hline \hline
    Epoch & Mid Time [BJD$_{TDB}$] & Reference \\ \hline
    -411.0 & 2456258.062876 $\pm$ 0.0002 & \citet{hellier_wasp96} \\
    201.0 & 2458354.320536 $\pm$ 0.000885 & This Work (T)\\
    202.0 & 2458357.744414 $\pm$ 0.000859 & This Work (T) \\
    203.0 & 2458361.170328 $\pm$ 0.000851 & This Work (T) \\
    204.0 & 2458364.594941 $\pm$ 0.000814 & This Work (T) \\
    206.0 & 2458371.446218 $\pm$ 0.000895 & This Work (T) \\
    207.0 & 2458374.871373 $\pm$ 0.000854 & This Work (T) \\
    208.0 & 2458378.296724 $\pm$ 0.000876 & This Work (T) \\
    235.0 & 2458470.777446 $\pm$ 0.001133 & This Work (H) \\
    238.0 & 2458481.054632 $\pm$ 0.000184 & This Work (H) \\
    325.0 & 2458779.050000 $\pm$ 0.001200 & This Work (S) \\
    327.0 & 2458785.900800 $\pm$ 0.000600 & This Work (S) \\ \hline 
    \multicolumn{3}{c}{T: TESS, H: Hubble, S: Spitzer}\\ \hline \hline
    \end{tabular}
    \caption{Transit mid times used to refine the ephemeris of planets from this study. Data which was originally in HJD time format was converted using the tool from \citet{Eastman}.}
    \label{tab:mid_times}
\end{table}

\begin{figure}
    \centering
    \includegraphics[width = \columnwidth]{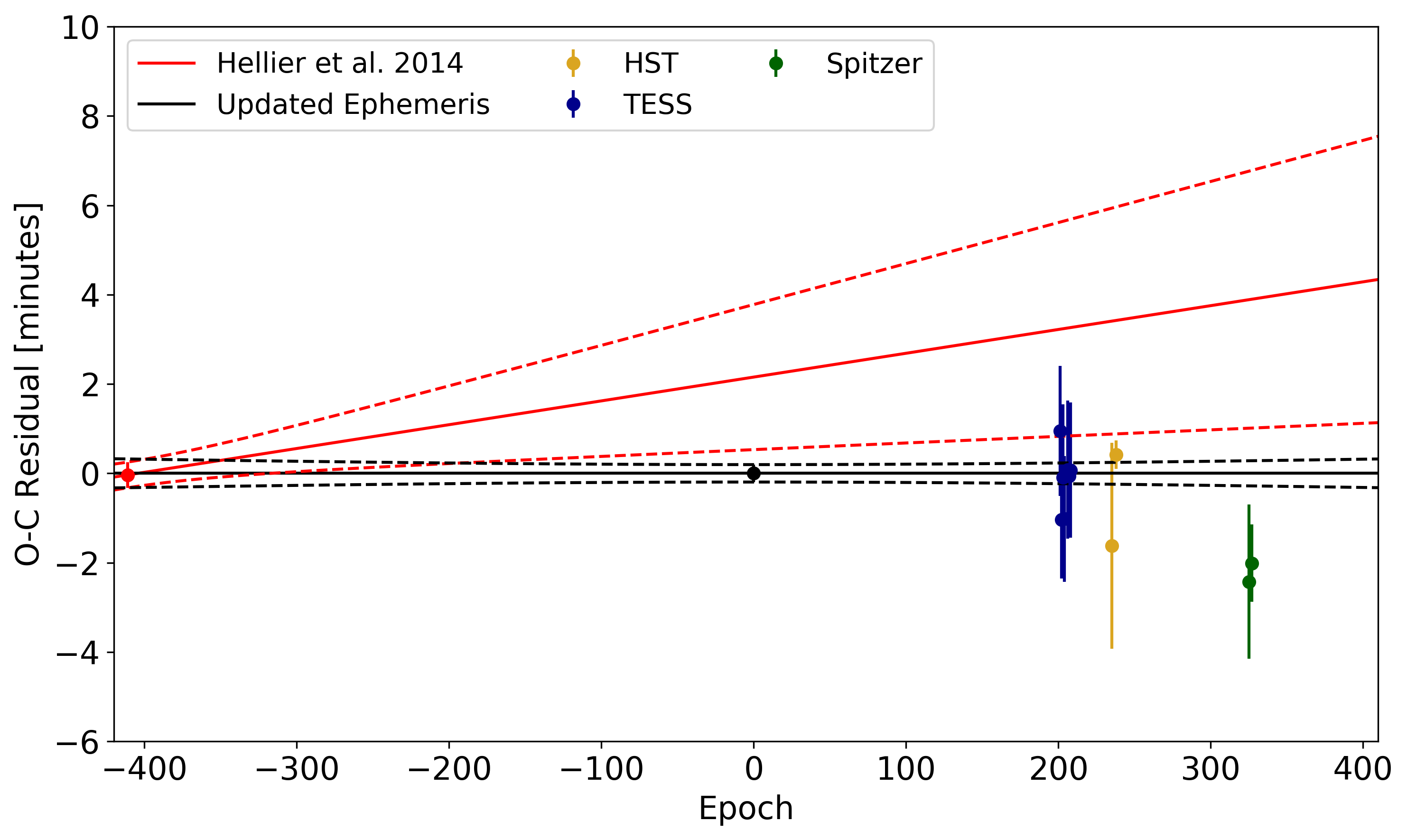}
    \caption{Observed minus calculated (O-C) mid-transit times for WASP-96\,b. Transit mid time measurements from this work are shown in gold (HST), blue (TESS) and green (Spitzer), while the $T_0$ value from \citet{hellier_wasp96} is in red. The black line denotes the new ephemeris of this work with the dashed lines showing the associated 1$\sigma$ uncertainties and the black data point indicating the updated T$_0$. For comparison, the previous literature ephemeris and their 1$\sigma$ uncertainties are given in red.}
    \label{fig:ephm_refine}
\end{figure}

We found that the observed HST and TESS transits were just outside the 1$\sigma$ literature ephemeris. Hence, we refined the period and reference mid transit time using the original ephemeris from \cite{hellier_wasp96}, and the new data analysed. We determined the ephemeris of WASP-96\,b to be $P = 3.42525650 \pm 0.00000043 $ days and $T_0$ = 2457665.84332 $\pm$ 0.00014 BJD$_{TDB}$ where $P$ is the planet's period, $T_0$ is the reference mid-time of the transit and BJD$_{TDB}$ is the barycentric julian date in the barycentric dynamical frame. Our derived period is 0.32 s shorter than that from \cite{hellier_wasp96} and we improved the precision of the period by a factor of 6, thus reducing the current uncertainty on the transit time. The observed minus calculated plots are given in Figure \ref{fig:ephm_refine} and all transit mid times used for the fitting are listed in Table \ref{tab:mid_times}. We note that the Spitzer points are both seemingly poor fits to the trend while the second HST observation (G141 grism) gives extremely tight bounds on the mid time despite the gaps within the light curve. We attempted an ephemeris fit without the HST observations and found little change in the period. TESS will soon re-observe WASP-96\,b which will allow for further refinement of its period. The mid times have been uploaded to ExoClock\footnote{\url{https://www.exoclock.space}}, an initiative to ensure transiting planets are regularly followed-up, keeping their ephemeris up-to-date for the ESA Ariel mission \citep{tinetti_ariel, edwards_ariel}.

\subsection{WASP-96\,b in Context}

\begin{figure}
    \centering
    \includegraphics[width = \columnwidth]{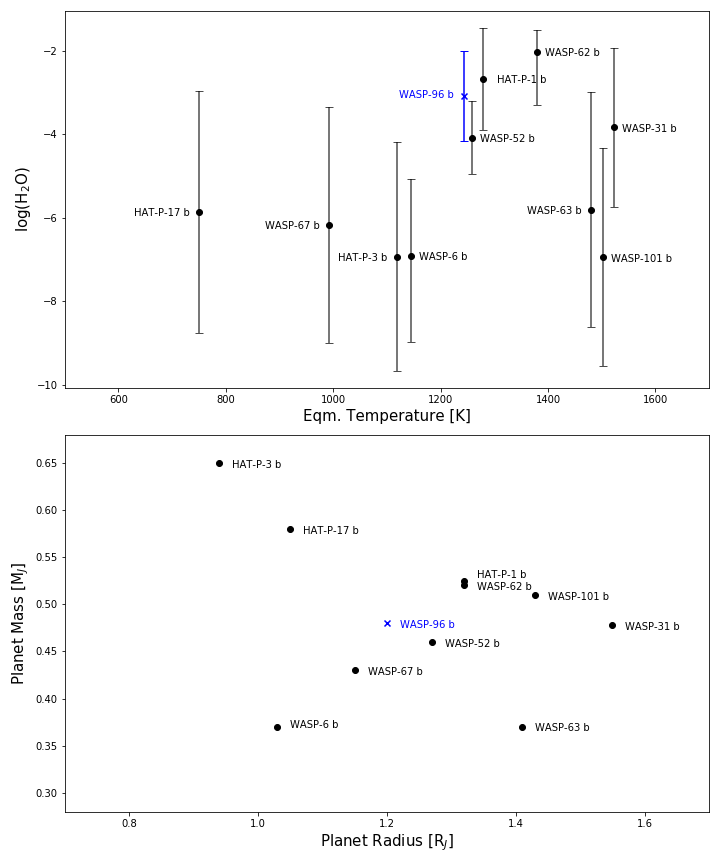}
    \caption{Comparing WASP-96\,b with planets with a similar size and mass. Top: Retrieved water log abundance of each planet against their respective equilibrium temperature. The water content of WASP-96\,b, WASP-52\,b and HAT-P-1\,b are similar, as are their equilibrium temperatures. Bottom: Mass-Radius plot of exoplanets within $\pm$ 0.5 R$_J$ and $\pm$ 0.2 M$_J$ of WASP-96\,b's radius and mass.}
    \label{fig:W96_compare}
\end{figure}

Water appears to be ubiquitous in exoplanetary atmospheres. To understand the distribution in the abundance of this molecule, we compared the retrieved water (log) abundance (log(H$_2$O) of exoplanets with similar sizes ($\pm$ 0.5 R$_J$) and masses ($\pm$ 0.2 M$_J$) against their respective equilibrium temperature (see top panel of Figure \ref{fig:W96_compare}) using data from \citet{tsiaras_30planets,pinhas,skaf2020}. Whilst these trends are not statistically significant yet, at temperatures of 1200-1400 K, planets appear to generally have a high water abundance. These planets are also closest in terms of size and mass to WASP-96\,b (lower panel of Figure \ref{fig:W96_compare}). An HST WFC3 study of Kepler-51\,b \& d, which have radii of 0.61 R$_J$ and 0.84 R$_J$ respectively, uncovered flat spectra for both, with no discernible atmospheric features \citep{libby_k51}. However, these planets have noticeably cooler temperatures than WASP-96\,b (400-550 K) and very low densities ($<$ 0.07 $g/cm^3$). While the addition of our detection adds weight towards the tendency of having water rich atmosphere for this class of hot planets, more objects of a similar class are needed in order to statistically verify this claim.

\section{Conclusions}

WASP-96 b is one of many planets to have been observed by both space and ground-based instruments. Each instrument is sensitive to different chemical species and the combined wavelength spans from optical to near-infrared. In this investigation we de-trended light-curves obtained from WFC3/HST and detected strong evidence for the presence of water, along with ruling out a large abundance of NH$_3$, in the atmosphere of this hot-Jupiter. 

As we tried to combine our data with observations from \citet{Nikolov2018}, we observed a large offset (1197 $^{+134}_{-101}$ ppm) from the combined transmission spectrum. The inconsistency between them rendered any retrieval to be impossible without any correction. We thus fit for an offset parameter during our retrieval on the combined observation, in an attempt to correct for the discrepancy. The combined retrieval shows a consistent water and sodium abundance with analyses on individual instruments. It was also able to retrieve a consistent temperature with the equilibrium temperature of the planet. 

Despite having seemingly better aligned spectrum and better constrained result after the correction was made, we would like to point out that such correction does not make the two observations compatible, and in fact there is no theoretical base for such correction, and therefore any conclusion should be taken with care. 

The case of WASP-96 b served as an alarming example that compatibility between different instruments, be it space-based or ground-based, should not be taken for granted. Seemingly consistent observations do not necessarily mean they \textit{are} consistent. It is especially true when the observations do not overlap in wavelength.

The difficulty in confirming the compatibility between instruments will be mitigated for with the next generation instrumentation such as JWST, Ariel \citep{tinetti_ariel} or Twinkle \citep{edwards_exo}. Their broad, simultaneous wavelength coverage will provide continuous coverage from optical to far-infrared at unprecedented resolution and SNR, which could resolve the trouble of having to combine observations in exchange for broader wavelength coverage.

\vspace{5mm}
\textbf{Acknowledgements:} This project has received funding from the European Research Council (ERC) under the European Union's Horizon 2020 research and innovation programme (grant agreement No 758892, ExoAI) and under the European Union's Seventh Framework Programme (FP7/2007-2013)/ ERC grant agreement numbers 617119 (ExoLights). Furthermore, we acknowledge funding by the ASI grant n. 2018.22.HH.O and by the Science and Technology Funding Council (STFC) grants: ST/K502406/1, ST/P000282/1, ST/P002153/1, ST/T001836/1 and ST/S002634/1. 

\vspace{5mm}
\textbf{Software:} Iraclis \citep{tsiaras_hd209}, TauREx3 \citep{al-refaie_taurex3}, pylightcurve \citep{tsiaras_plc}, ExoTETHyS \citep{morello_exotethys}, Astropy \citep{astropy}, h5py \citep{hdf5_collette}, emcee \citep{emcee}, Matplotlib \citep{Hunter_matplotlib}, Multinest \citep{Feroz_multinest}, corner \citep{corner}, Pandas \citep{mckinney_pandas}, Numpy \citep{oliphant_numpy}, SciPy \citep{scipy}.

\vspace{5mm}
%\textbf{Data:} This work is based upon observations with the NASA/ESA Hubble Space Telescope, obtained at the Space Telescope Science Institute (STScI) operated by AURA, Inc. The publicly available HST observations presented here were taken as part of proposal 15469, led by Nikolay Nikolov. These were obtained from the Hubble Archive which is part of the Mikulski Archive for Space Telescopes. Additionally we have used data from Spitzer.. \citet{nikolov_spitzer}

\textbf{Data:} This work is based upon observations with the NASA/ESA Hubble Space Telescope, obtained at the Space Telescope Science Institute (STScI) operated by AURA, Inc. The publicly available HST observations presented here were taken as part of proposal 15469, led by Nikolay Nikolov. We are grateful to all those involved in the creation of the proposals which led to these valuable HST and Spitzer data sets being made available to the community: Nikolay Nikolov, Gilda E. Ballester, Aarynn Carter, Drake Deming, Ben Drummond, Jonathan Fortney, Neale Gibson, Jayesh Goyal, Tiffany Kataria, Nathan J. Mayne, Thomas Mikal-Evans, David K. Sing, Jessica Spake, and Hannah Wakeford. These were obtained from the Hubble Archive which is part of the Mikulski Archive for Space Telescopes. This paper also includes data collected by the TESS mission, which are publicly available from the Mikulski Archive for Space Telescopes (MAST) and produced by the Science Processing Operations Center (SPOC) at NASA Ames Research Center \citep{jenkins_tess}. This research effort made use of systematic error-corrected (PDC-SAP) photometry \citep{smith_pdc,stumpe_pdc1,stumpe_pdc2}. Funding for the TESS mission is provided by NASA’s Science Mission directorate. This work is also based in part on observations made with the Spitzer Space Telescope, which is operated by the Jet Propulsion Laboratory, California Institute of Technology, under a contract with NASA. These observations were taken as part of proposal 14255, led by Nikolay Nikolov \citep{nikolov_spitzer}.

%We are thankful to those who operate this archive, the public nature of which increases scientific productivity and accessibility \citep{peek2019}.

\begin{table*}
    \centering
    \resizebox{\textwidth}{!}{
    \begin{tabular}{|cccccc|ccccc|} \hline \hline 
 Wavelength $[\mu m$]  & Transit Depth [\%] & Error [\%] & Bandwidth [$\mu m$] & Instrument & & Wavelength [$\mu m$] & Transit Depth [\%] & Error [\%] & Bandwidth [$\mu m$] & Instrument\\  \hline \hline 

0.37565	&	1.4065	&	0.0349	&	0.05130	& VLT FORS2 &		&	0.77330	&	1.3685	&	0.0247	&	0.00800 & VLT FORS2\\
0.40530	&	1.3992	&	0.0417	&	0.00800	& VLT FORS2 &		&	0.78130	&	1.3750	&	0.0218	&	0.00800	& VLT FORS2\\
0.41330	&	1.3844	&	0.0380	&	0.00800	& VLT FORS2 &		&	0.78930	&	1.3739	&	0.0279	&	0.00800	& VLT FORS2\\
0.42130	&	1.4238	&	0.0386	&	0.00800	& VLT FORS2 &		&	0.79730	&	1.3628 	&	0.0264	&	0.00800	& VLT FORS2\\
0.42930	&	1.4019	&	0.0470	&	0.00800	& VLT FORS2 &		&	0.81250	&	1.3532	&	0.0179	&	0.00250 & HST G102\\
0.43730	&	1.3705	&	0.0369	&	0.00800	& VLT FORS2 &		&	0.83750	&	1.3806	&	0.0239	&	0.00250 & HST G102\\
0.44530	&	1.4125	&	0.0297	&	0.00800	& VLT FORS2 &		&	0.86250	&	1.3792	&	0.0220	&	0.00250	& HST G102\\
0.45330	&	1.4035	&	0.0282	&	0.00800	& VLT FORS2 &		&	0.88750	&	1.3829	&	0.0313	&	0.00250	& HST G102\\
0.46130	&	1.4026	&	0.0287	&	0.00800	& VLT FORS2 &		&	0.91250	&	1.3800	&	0.0311	&	0.00250	& HST G102\\
0.46930	&	1.3796	&	0.0264	&	0.00800	& VLT FORS2 &		&	0.93750	&	1.3996	&	0.0189	&	0.00250	& HST G102\\
0.47730	&	1.3617	&	0.0226	&	0.00800	& VLT FORS2 &		&	0.96250	&	1.3574	&	0.0205	&	0.00250	& HST G102\\
0.48530	&	1.3712	&	0.0294	&	0.00800	& VLT FORS2 &		&	0.98750	&	1.3804	&	0.0215	&	0.00250	& HST G102\\
0.49330	&	1.3976	&	0.0238	&	0.00800	& VLT FORS2 &		&	1.01250	&	1.3789	&	0.0199	&	0.00250	& HST G102\\
0.50130	&	1.4001	&	0.0316	&	0.00800	& VLT FORS2 &		&	1.03750	&	1.4053	&	0.0216	&	0.00250	& HST G102\\
0.50930	&	1.3716	&	0.0254	&	0.00800	& VLT FORS2 &		&	1.06250	&	1.3796	&	0.0234	&	0.00250	& HST G102\\
0.51730	&	1.3698	&	0.0229	&	0.00800	& VLT FORS2 &		&	1.08750	&	1.3965	&	0.0140	&	0.00250	& HST G102\\
0.52530	&	1.3696	&	0.0217	&	0.00800	& VLT FORS2 &		&	1.11250	&	1.4212	&	0.0111	&	0.00250	& HST G102\\
0.53330	&	1.3985	&	0.0204	&	0.00800	& VLT FORS2 &		&	1.12625	&	1.3899	&	0.0162	&	0.02190	& HST G141\\
0.54130	&	1.3805	&	0.0216	&	0.00800	& VLT FORS2 &		&	1.14775	&	1.4110	&	0.0200	&	0.02110	& HST G141\\
0.54930	&	1.3796	&	0.0202	&	0.00800	& VLT FORS2 &		&	1.16860	&	1.4116	&	0.0210	&	0.02060	& HST G141\\
0.55730	&	1.3707	&	0.0231	&	0.00800	& VLT FORS2 &		&	1.18880	&	1.3875	&	0.0181	&	0.01980	& HST G141\\
0.56530	&	1.4208	&	0.0219	&	0.00800	& VLT FORS2 &		&	1.20835	&	1.3664	&	0.0217	&	0.01930	& HST G141\\
0.57330	&	1.4111	&	0.0218	&	0.00800	& VLT FORS2 &		&	1.22750	&	1.3811	&	0.0210	&	0.01900	& HST G141\\
0.58130	&	1.4468	&	0.0196	&	0.00800	& VLT FORS2 &		&	1.24645	&	1.3730	&	0.0152	&	0.01890	& HST G141\\
0.58930	&	1.4542	&	0.0215	&	0.00800	& VLT FORS2 &		&	1.26550	&	1.3809	&	0.0185	&	0.01920	& HST G141\\
0.59730	&	1.4456	&	0.0200	&	0.00800	& VLT FORS2 &		&	1.28475	&	1.4164	&	0.0185	&	0.01930	& HST G141\\
0.60530	&	1.4361	&	0.0200	&	0.00800	& VLT FORS2 &		&	1.30380	&	1.3804	&	0.0191	&	0.01880 & HST G141\\
0.61330	&	1.4217	&	0.0187	&	0.00800	& VLT FORS2 &		&	1.32260	&	1.4031	&	0.0229	&	0.01880	& HST G141\\
0.62130	&	1.4084	&	0.0221	&	0.00800	& VLT FORS2 &		&	1.34145	&	1.3927	&	0.0180	&	0.01890	& HST G141\\
0.62930	&	1.4456	&	0.0231	&	0.00800	& VLT FORS2 &		&	1.36050	&	1.4132	&	0.0170	&	0.01920	& HST G141\\
0.63730	&	1.4015	&	0.0220	&	0.00800	& VLT FORS2 &		&	1.38005	&	1.4429	&	0.0238	&	0.01990	& HST G141\\
0.64530	&	1.4111	&	0.0209	&	0.00800	& VLT FORS2 &		&	1.40000	&	1.4585	&	0.0241	&	0.02000 & HST G141\\
0.65330	&	1.4236	&	0.0231	&	0.00800	& VLT FORS2 &		&	1.42015	&	1.4273	&	0.0206	&	0.02030 & HST G141\\
0.66130	&	1.3621	&	0.0201	&	0.00800	& VLT FORS2 &		&	1.44060	&	1.4345	&	0.0191	&	0.02060	& HST G141\\
0.66930	&	1.4187	&	0.0187	&	0.00800	& VLT FORS2 &		&	1.46150	&	1.3797	&	0.0227	&	0.02120	& HST G141\\
0.67730	&	1.3980	&	0.0268	&	0.00800	& VLT FORS2 &		&	1.48310	&	1.4007	&	0.0183	&	0.02200	& HST G141\\
0.68930	&	1.3408	&	0.0175	&	0.01600	& VLT FORS2 &		&	1.50530	&	1.4038	&	0.0218	&	0.02240	& HST G141\\
0.70130	&	1.3682	&	0.0204	&	0.00800	& VLT FORS2 &		&	1.52800	&	1.3887	&	0.0190	&	0.02300	& HST G141\\
0.70930	&	1.4074	&	0.0220	&	0.00800	& VLT FORS2 &		&	1.55155	&	1.3661	&	0.0154	&	0.02410	& HST G141\\
0.71730	&	1.3948	&	0.0194	&	0.00800	& VLT FORS2 &		&	1.57625	&	1.3945	&	0.0170	&	0.02530	& HST G141\\
0.72530	&	1.4029	&	0.0222	&	0.00800	& VLT FORS2 &		&	1.60210	&	1.3796	&	0.0235	&	0.02640	& HST G141\\
0.73330	&	1.4280	&	0.0218	&	0.00800	& VLT FORS2 &		&	1.62945	&	1.3791	&	0.0217	&	0.02830	& HST G141\\
0.74130	&	1.3787	&	0.0184	&	0.00800	& VLT FORS2 &		&	0.8     &   1.3803  &   0.0346  &   0.4 & TESS\\
0.74930	&	1.3896	&	0.0219	&	0.00800	& VLT FORS2 &		&	3.6     &   1.4796  &   0.0860  &   0.75   & Spitzer IRAC\\
0.76130	&	1.4052	&	0.0172	&	0.01600	& VLT FORS2	&       &	4.5     &   1.4861  &   0.0538  &   1.015  & Spitzer IRAC\\ \hline \hline	
    \end{tabular}
    }
    \caption{The transit depths derived here for TESS, HST and Spitzer along with the VLT data, post offset correction.}
    \label{tab:spec_data}
\end{table*}

\clearpage
\appendix
\section{HST spectrum with more complicated model}
\label{fig:HST_complex_posterior}
\begin{figure*}
    \centering
    \includegraphics[width = 0.75\textwidth]{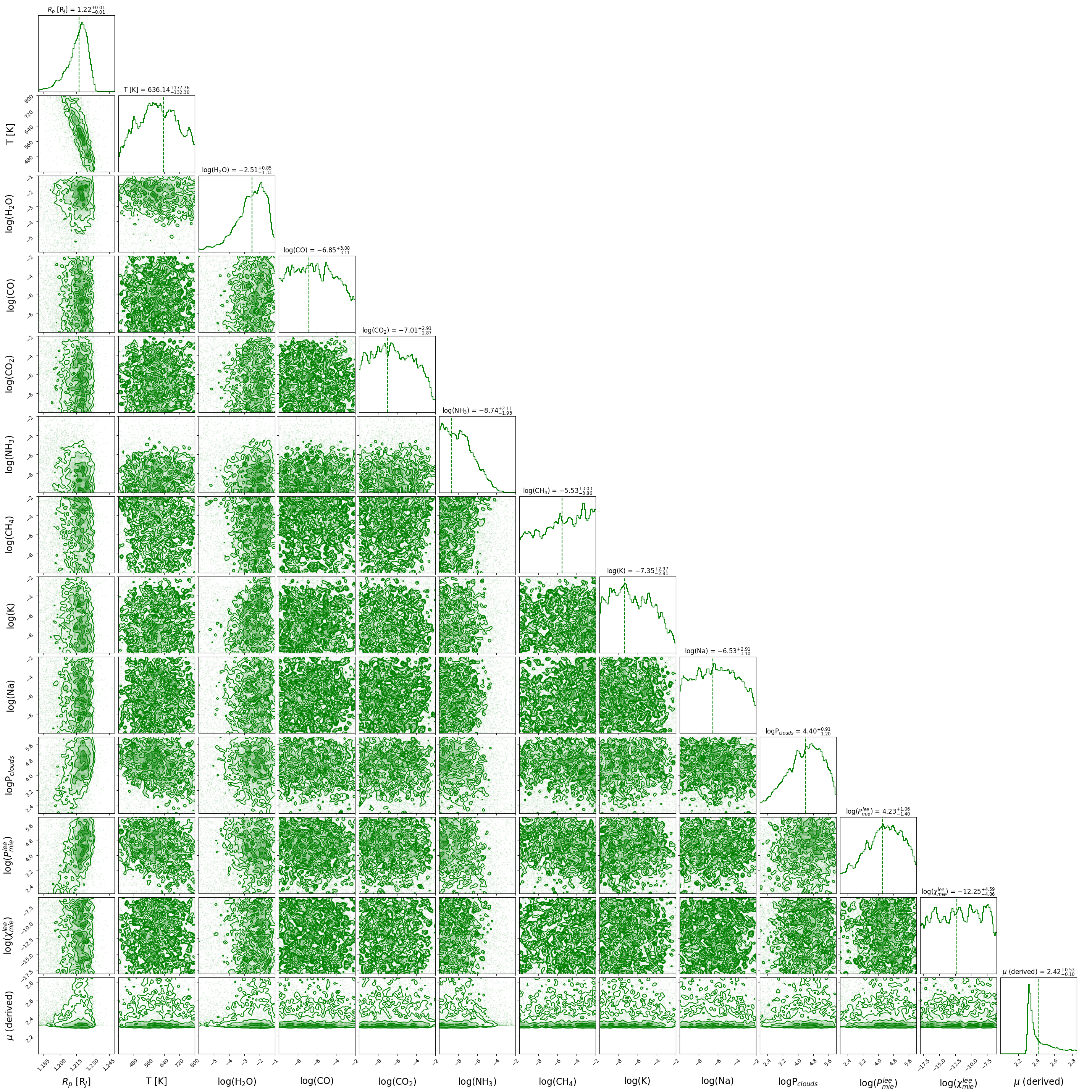}
    \caption{Posteriors distribution of different atmospheric parameters retrieved using HST data with more complex atmospheric models. }
    \label{fig:hst_complex}
\end{figure*}

For better comparison with our results from VLT + HST observations (Figure \ref{fig:offset_posteriors} and Figure \ref{fig:W96_compare}), we applied the same atmospheric setup as VLT+HST retrieval to HST observations so as to understand the outcome from a more complicated model (See Figure \ref{fig:hst_complex}). The result of the posterior distribution shows a similar outcome to our simpler model. The model was not able to constrain additional parameters, which is expected given the limited wavelength range from the instrument, hence we opted for simpler model to present in the main text.
\clearpage
\bibliographystyle{aasjournal}
\bibliography{main}

\end{document}